\documentclass[a4paper,12pt]{article}
\usepackage{amsfonts,amsthm,amsmath,amssymb,graphicx,hyperref,color,youngtab}

\newcommand{\be}{\begin{equation}}
\newcommand{\bea}{\begin{eqnarray}}

\newcommand{\ee}{\end{equation}}
\newcommand{\eea}{\end{eqnarray}}

\begin{document}

\makeatletter
\@addtoreset{equation}{section}
\makeatother
\renewcommand{\theequation}{\thesection.\arabic{equation}}
\vspace{1.8truecm}

{\LARGE{\centerline{\bf Chaos in the Fishnet}}}  

\vskip.5cm 

\thispagestyle{empty} 
\centerline{ {\large\bf Robert de Mello Koch$^{a,b,}$\footnote{{\tt robert@neo.phys.wits.ac.za}}, W. LiMing$^{a,}$\footnote{{\tt wliming@scnu.edu.cn}}}}
\centerline{{\large\bf Hendrik J.R. Van Zyl${}^{b,}$\footnote{ {\tt hjrvanzyl@gmail.com}}
and Jo\~ao P. Rodrigues$^{b,}$\footnote{{\tt Joao.Rodrigues@wits.ac.za}} }}

\vspace{.4cm}
\centerline{{\it ${}^{a}$ School of Physics and Telecommunication Engineering},}
\centerline{{ \it South China Normal University, Guangzhou 510006, China}}

\vspace{.4cm}
\centerline{{\it ${}^{b}$ National Institute for Theoretical Physics,}}
\centerline{{\it School of Physics and Mandelstam Institute for Theoretical Physics,}}
\centerline{{\it University of the Witwatersrand, Wits, 2050, } }
\centerline{{\it South Africa } }

\vspace{1truecm}

\thispagestyle{empty}

\centerline{\bf ABSTRACT}

\vskip.2cm 

We consider the computation of out-of-time-ordered correlators (OTOCs) in the fishnet theories, with a mass term added.
These fields theories are not unitary.
We compute the growth exponent, in the planar limit, at any value of the coupling and show that the model exhibits chaos.
At strong coupling the growth exponent violates the Maldacena-Shenker-Stanford bound. 
We also consider the mass deformed versions of the six dimensional honeycomb theories, which can also be solved in the planar limit.
The honeycomb theory shows a very similar behavior to that exhibited by the fishnet theory.

\setcounter{page}{0}
\setcounter{tocdepth}{2}
\newpage
\tableofcontents
\setcounter{footnote}{0}
\linespread{1.1}
\parskip 4pt

{}~
{}~

\section{Introduction}

Chaos in many-body quantum systems can be probed using non-time-ordered four point functions\cite{LO,Almheiri:2013hfa,Shenker:2013pqa,Shenker:2013yza,Roberts:2014isa,Roberts:2014ifa,Stanford:2015owe}. 
In this language, the butterfly effect is the statement that for rather general operators $W$ and  $V$, 
the thermal expectation value of the square of the commutator
\bea
c(t) = \langle[W(t), V ][W(t), V ]^\dagger\rangle_\beta
\eea
becomes large at late time.
For simple operators in large $N$ systems, there is a long period of exponential growth\cite{Roberts:2014isa} and generically we expect
\bea
  c(t) \propto {1\over N^2}e^{\lambda_L t}
\eea
The rate $\lambda_L$ defined by this exponential quantifies the strength of chaos and remarkably, it is governed by a bound $\lambda_L\le {2\pi\over\beta}$\cite{Maldacena:2015waa}.

The exquisite paper \cite{Stanford:2015owe} evaluated $\lambda_L$ in a weakly coupled large $N$ quantum field theory, at finite temperature.
Our study is heavily influenced by \cite{Stanford:2015owe} and we are repeating the same analysis for a much simpler class of theories.
For this reason it is useful to review salient parts of \cite{Stanford:2015owe}. 
The model considered is of a single hermitian matrix $\Phi_{ab}(\vec x,t)$ of mass $m$ interacting through a ${\rm Tr}(\Phi^4)$ self coupling.
The observable considered is a color and spatially-averaged version of the squared commutator (we indicate the color sums explicitly and use ${\rm Tr}$ for the thermal average computed with thermal density matrix $\rho$ at inverse temperature $\beta$)
\bea
C(t) = {1\over N^4}\sum_{aba'b'}\int d^3 x{\rm Tr}\left(
\sqrt{\rho}\left[\Phi_{ab}(t, \vec{x}), \Phi_{a',b'}(0,\vec{0})\right]
\sqrt{\rho}\left[\Phi_{ab}(t, \vec{x}), \Phi_{a'b'}(0,\vec{0})\right]^\dagger\right)
\eea
By splitting $\rho$ into two square root factors, we place the two commutators on opposite sides of the thermal circle, which nicely regulates some divergences.
The regularization dependence of $\lambda_L$ has been discussed in \cite{Romero-Bermudez:2019vej}. 
The computation simplifies at large $N$ since only the planar diagrams contribute.
Expanding, the two commutators gives four terms with each computed by a particular analytic continuation of the Euclidean correlator. 
Each term can also be represented with path integral contours in complex time, where some real-time folds are appended to the Euclidean thermal circle. 
For each term we generate the perturbation series by expanding the interaction vertex, integrating each vertex along the contour and then applying Wick's Theorem with contour-ordered propagators\cite{Stanford:2015owe}.
In \cite{Stanford:2015owe} the region of integration for the interaction vertices is restricted to the real-time folds. 
The rationale for this restriction is that the integral over the thermal circle corrects the thermal state.
Although these corrections are important for getting the exact $C(t)$, the claim is that they do not affect the spectrum of
growth exponents. 
This is intuitively convincing and further, these two simplifications are valid at any coupling.
We will employ the same simplifications in our study.

To proceed \cite{Stanford:2015owe} makes one more approximation. 
To compute $\lambda_L$ to leading order in the coupling, only the fastest-growing function of time at each order is retained.
This simplification is dramatic: it restricts the diagrams that are to be summed to the dressed ladder diagrams. 
The rate of growth of the sum of ladder diagrams becomes the problem of finding the largest eigenvalue of a one-dimensional integral equation.
This can be diagonalized numerically.
The resulting $\lambda_L$ can be determined explicitly.
One point that we have taken note of, is that $\lambda_L$ is proportional to $m^{-1}$ indicating that the important degrees of freedom are the highly populated, frequently colliding low energy quanta with $E \sim m$.
The dominant effect does not come from the thermal scale quanta as one might have thought.
For the systems we study we find a divergent $\lambda_L$ as $m\to 0$ suggesting a similar behavior.

Our goal is to carry out the same analysis, but for the fishnet model deformed with a mass term.
We will also ultimately need to resort to numerical methods and the mass deformation avoids difficulties from the $m^{-1}$ dependence of $\lambda_L$. 
The fishnet family of models, proposed in \cite{Gurdogan:2015csr} are non-unitary, non-supersymmetric CFTs obtained by taking a special double scaling limit of $\gamma$-deformed ${\cal N}=4$ super Yang-Mills theory.
This limit combines a weak coupling limit (the 't Hooft coupling is sent to zero) with an infinite imaginary twist, $\gamma_j\to i\infty$ leaving the theory with three finite effective couplings $\xi_j=\lambda e^{−i\gamma_j/2}$.
The limit decouples the gauge fields and the gaugino so that the model has three complex scalars and three complex fermions with a certain chiral interaction.
This chiral interaction significantly reduces the number of Feynman diagrams that can be drawn, so that at large $N$ there is often only a single diagram at a given order in the 't Hooft coupling!
In this way the double scaling limit significantly simplifies computations of interesting physical properties\cite{Gromov:2018hut,Chicherin:2017cns,Chicherin:2017frs,Korchemsky:2018hnb}.
The fishnet was further studied in \cite{Caetano:2016ydc} using asymptotic Bethe ansatz methods.
In all of these results the conformal symmetry plays a crucial role.
We have spoiled this symmetry with our mass deformation.
Nonetheless, the deformed fishnet continues to provide a beautiful model because there are still very few Feynman diagrams that contribute.
Indeed, we will again find that the computation of the OTOC reduces to summing ladders which occurs because of the specific structure of the fishnet vertex and not because we assume weak coupling.
In this way, the fishnet provides a simple solvable toy model that exhibits chaotic dynamics.
This is discussed in detail in section \ref{FishnetOTOC}.
Our numerical results for $\lambda_L$, developed in section \ref{NumbersFishnet}, show the behavior of $\lambda_L$ as a function of mass for $0.001<m\beta<0.05$. For example, we have
\bea
\lambda_L\approx 2{\lambda^2\over\beta}
\eea
at $m\beta =0.01$.
Consequently, as the 't Hooft coupling is increased $\lambda_L$ will violate the bound of \cite{Maldacena:2015waa}.
Recall that the fishnet theory is not unitary.
The authors of \cite{Maldacena:2015waa} suggest that unitarity, analyticity and causality are the crucial assumptions 
necessary to prove the bound. 
They used Hamiltonian systems for which unitarity and causality are manifest.
Our result supports the idea that unitarity is an important ingredient in the proof of \cite{Maldacena:2015waa}.

The fishnet theory has close cousins in both 3 and 6 dimensions. 
We consider the six dimensional theory in section \ref{hexOTOC}, again reducing the problem of determining $\lambda_L$ to finding the largest eigenvalue of a one-dimensional integral equation. We do this numerically in section \ref{HexNumbers}.
Our numerical results again show that the chaos bound is violated at strong coupling.

\section{Fishnet OTOC}\label{FishnetOTOC}

The model that we study is described by the Lagrangian density
\bea
   {\cal L}=N{\rm Tr}\left( \partial^\mu X^\dagger\partial_\mu X
                                    +\partial^\mu Y^\dagger\partial_\mu Y
                                    -m^2 XX^\dagger -m^2 YY^\dagger - \lambda X^\dagger Y^\dagger X Y\right)\label{FishnetLag}
\eea
The fishnet CFT would have 3 complex adjoint scalars.
For our purposes it is enough to study a model with two complex scalars so, for simplicity, we will do this.
The model is not unitary because the interaction term is not hermitian.
\begin{figure}[ht]%
\begin{center}
\includegraphics[width=0.5\columnwidth]{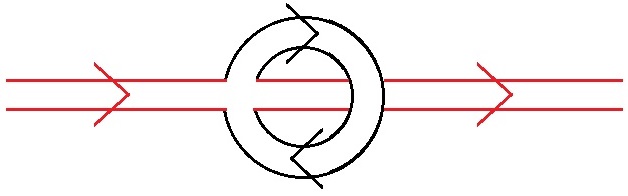}%
\caption{The one loop mass correction is a non-planar graph. There are no planar corrections to the two point function.}%
\label{NoCorrect2Pnt}
\end{center}
\end{figure}
Our first observation is that there are no planar diagrams that correct the two point function.
For example, the one loop mass correction and two loop wave function correction are both non-planar graphs, thanks to the specific form of the quartic fishnet interaction. 
See Figure \ref{NoCorrect2Pnt} where the one loop mass correction graph is shown. 
Thus, at large $N$ and at any value of the coupling, the two point functions are given by their free field values.
As explained in \cite{Stanford:2015owe} both the retarded Green's function and the Wightman functions are needed.
These are given by
\bea
G_R(k)&=&{i\over 2\omega_{\vec{k}}}\left({1\over k^0-\omega_{\vec{k}}+i\epsilon}
-{1\over k^0+\omega_{\vec{k}}+i\epsilon}\right)\cr\cr
\tilde G(k)&=&\sum_{s=\pm}{\pi\delta (k^0-s\omega_{\vec{k}})\over 2\omega_{\vec{k}}
\sinh {\beta\omega_{\vec{k}}\over 2}}
\eea
The rungs of the ladder are given by Wightman correlators while the side rails are retarded propagators.
This structure is not too difficult to appreciate: due to the presence of the density matrices in (\ref{CrrWeWant}),
we are working on a pair of folded time contours.
The interaction vertices should be integrated over both sides of each fold. 
Each side comes with a different sign and a different ordering of the operators so that the side rails of the ladder diagrams 
turn into retarded propagators.
The rungs remain Wightman correlators.

Next consider the large $N$ limit of the four point function\footnote{For a beautiful computation of the four point function in the planar limit of the conformal model, see \cite{Kazakov:2018gcy}.}.
The only diagrams that contribute are the ladder diagrams and, as we have just explained, none of the rungs or side rails are dressed at large $N$.
Consequently, the sum of ladder diagrams satisfies the integral equation 
\bea
f(\omega,p)&=&-G_R(p)G_R(\omega-p)\left[1+\int {d^4 k\over (2\pi)^4} R(k-p)f(\omega,k)\right]\label{Ieqb}
\eea
where the kernel of the integral equation is\footnote{This integral can be performed exactly. See \cite{Stanford:2015owe,Jeon:1994if} for the details.}
\bea
R(k)&=&\lambda^2\int {d^4 p\over (2\pi)^4}\tilde G(k/2-p)\tilde G(k/2+p)\cr
&=&{\lambda^2\over 8\pi\beta |\vec k|\sinh {|k^0|\beta\over 2}}\left[
\theta (k^2-4m^2)\log {\sinh x_+\over\sinh x_-}
+\theta (-k^2)
\log{1-e^{-2x_+}\over 1-e^{2x_-}}
\right]\cr\cr\cr
x_\pm&=&{\beta\over 4}\left(
|k^0|\pm |\vec k|\sqrt{1+{4m^2\over \vec k\cdot\vec k-(k^0)^2}}
\right)\qquad\qquad k^2=(k^0)^2-\vec k\cdot\vec k
\eea
and where in (\ref{Ieqb}) the function $f(\omega,p)$ is defined by the Fourier transform of the double commutator
\bea
f(\omega,p)=-\sum_{aba'b'}\int d^4 x {e^{-i\omega t+i\vec{p}\cdot\vec{x}}\over N^2}{\rm Tr}\left(
\sqrt{\rho}\left[X_{ab}^\dagger (t, \vec{x}), X_{a',b'}(0,\vec{0})\right]
\sqrt{\rho}\left[X_{ab}^\dagger (t, \vec{x}), X_{a'b'}(0,\vec{0})\right]\right)\cr
\label{CrrWeWant}
\eea
The integral equation (\ref{Ieqb}) is exact at large $N$, not restricted to weak coupling.
This integral equation is very similar to the equation obtained in \cite{Stanford:2015owe}.
The only difference is that in \cite{Stanford:2015owe} the two point function is dressed with a two loop self energy. 
The imaginary part of this self energy leads to an exponential decay of the two point function reflecting the fact that
collisions with thermal excitations can knock the particle into a different momentum state so that it has a finite lifetime. 
As we explained above, this process vanishes at large $N$, thanks to the form of the fishnet interaction.
This correction drives thermalization, so that the planar fishnet theory does not thermalize.
For the exponential growth that we are interested in, we need the late time behavior of the correlator.
This is determined by a double pole whose contribution is isolated by replacing\cite{Stanford:2015owe}
\bea
G_R(p) G_R (\omega-p)\to 
-{\pi i\over 2\omega_{\vec{p}}^2}{\delta (p^0-\omega_{\vec p})+\delta (p^0+\omega_{\vec p})\over\omega +i\epsilon}
\label{latetime}
\eea
Now, making the ansatz 
\bea
  f(\omega ,p) = f(\omega,\vec{p})\delta (p_0^2-\omega_{\vec{p}}^2)\label{onshell}
\eea
and (after Fourier transforming $\omega$) $f(t,p)=e^{\lambda_L t}f(p)$ we find
\bea
\lambda_L f(p)&=&\int {d^3 k\over (2\pi)^3}{1\over 4\omega_{\vec k}\omega_{\vec{p}}}\left(R(\gamma_+)
+R(\gamma_-)\right)f(k)\cr\cr
\gamma_\pm &=& (\omega_{\vec{k}}\pm\omega_{\vec{p}},\vec{k}-\vec{p})\label{fnl}
\eea
We want to do the integral over $d^3 k$ in (\ref{fnl}).
From the explicit form of $R(\cdot)$ we know that $R(k_\pm)$ only depends on $|\vec p|$, $|\vec k|$ and $|\vec p-\vec k |$.
Changing to spherical coordinates, and using the symmetry of the problem, which implies that $f(p)=f(|\vec p|)$, we have
\bea
\lambda_L f(p)=\int_{-\infty}^{\infty} dk K(k) f(k)\label{eproblem}
\eea
where the kernel in the above equation is
\bea
K(k)= {k^2 \over 16 \pi^2\omega_{\vec k}\omega_{\vec p}}\int_0^{\pi} d\theta\,  \sin\theta
\left(R(\gamma_+)+R(\gamma_-)\right)\label{forkernel}
\eea
This defines an eigenvalue problem for the exponent $\lambda_L$ and is the equation that we study numerically 
in the next section. 

There is a final point that deserves discussion.
The Lagrangian (\ref{FishnetLag}) should be supplemented with extra double-trace interactions that are induced by 
quantum corrections\cite{Dymarsky:2005uh,Sieg:2016vap,Pomoni:2008de}. 
This double-trace renormalization is a leading effect at large $N$ and so can contribute in the planar limit.
The possible double trace terms are
\bea
{\cal L}_{dt}&=&\xi_1 \left({\rm Tr}(XX){\rm Tr}(X^\dagger X^\dagger)+{\rm Tr}(YY){\rm Tr}(Y^\dagger Y^\dagger)\right)
+\xi_2 {\rm Tr}(XY^\dagger){\rm Tr}(YX^\dagger)\cr
&&\qquad\qquad+\xi_3{\rm Tr}(XY){\rm Tr}(X^\dagger Y^\dagger)
\eea
For the four point correlator computed by $C(t)$ we only need to consider $\xi_1$.
The contribution of similar double trace interactions, to four point functions, was considered in \cite{Kazakov:2018gcy}.
In this work integral equations obeyed by the exact four point functions were obtained by introducing suitable 
graph-building operators.
The four point function is given by a sum of ladder operators, together with double trace interactions.
There are consequently two graph building operators: one that adds a rung to the ladder and one that adds a double trace interaction.
To start, the analysis ignores the contribution of the double trace interactions and sums the ladder diagrams by diagonalizing
the operator that adds a rung.
It is then a straight forward analysis to show that the operator that adds a double trace interaction annihilates the eigenfunctions of the operator that adds a rung and consequently the conclusion \cite{Kazakov:2018gcy} is that in the end the exact finite coupling solutions are not modified by the double trace interactions\footnote{We thank Volodya Kazakov for constructive correspondence on the double trace interactions.}.
 
For the computations considered in this paper the double trace interactions again play no role, but its for a rather different
reason. The contributions from the double trace interactions cancel when we take the sum of the terms to form
the square of the commutator. 
For a standard four-point function (without commutators) the double trace interactions make a non-zero contribution.

\section{Numerical Results}\label{NumbersFishnet}

To solve (\ref{eproblem}) we will perform the integral in (\ref{forkernel}) numerically and discretize $p$ and $k$ to obtain a matrix equation.
To perform the integral in (\ref{forkernel}) it is helpful to change integration variable from $\theta$ to $y=\sqrt{k^2+p^2-2kp\cos\theta}$.
It is convenient to change integration variables before we discretize (\ref{eproblem}). 
We set
\bea
k={3u\over u-1}\label{transk}
\eea
so that we map the range $0\le k<\infty$ to the range $0\le u<1$.
The eigenvalue problem for the exponent then becomes
\bea
\lambda_L f(u)=\int_0^1 \tilde K(u,v) f(v) dv
\eea
where, after a suitable redefinition of the eigenfunctions $f(u)$, the kernel $\tilde K(u,v)$ is symmetric 
so that we are guaranteed that it can be diagonalized.
The largest eigenvalue of the matrix obtained from $\tilde K(u,v)$ after we discretize is the growth exponent $\lambda_L$.

\begin{figure}[ht]%
\begin{center}
\includegraphics[width=0.8\columnwidth]{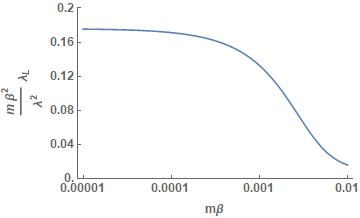}%
\caption{Behavior of the growth exponent $\lambda_L$ for the fishnet model.
}%
\label{growthexponent}
\end{center}
\end{figure}

Numerical convergence is rapid and the numerics converges after discretizing $u$ and $v$ uniformly with about 200 points.
Discretizing with a finer lattice does not change the value obtained for $\lambda_L$.
All eigenvalues are positive, in contrast to the results of \cite{Stanford:2015owe} where a continuum of negative eigenvalues were found.
The difference is that for the model of \cite{Stanford:2015owe} the retarded propagators receive a self energy correction, with an imaginary part at two loops.
This imaginary part leads to exponential decay of the two point correlation functions indicating a finite lifetime of single particle states.
This is because at finite temperature a particle can be knocked into a different momentum state through collision with a thermal excitation.
Our numerical results are given in Figure \ref{growthexponent}.

\section{Honeycomb OTOC}\label{hexOTOC}

\begin{figure}[ht]%
\begin{center}
\includegraphics[width=0.5\columnwidth]{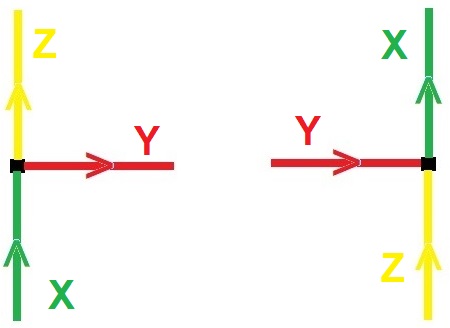}%
\caption{Vertices for the honeycomb model.}%
\label{vertices}
\end{center}
\end{figure}

We study a cubic theory in $d=6$ dimensions.
The coupling constant of the cubic theory is dimensionless so the massless theory is classically conformal.
Following our analysis of the fishnet theory, we will again add a mass term\footnote{For an interesting study of the massless theory with non-trivial results, the reader should consult \cite{Mamroud:2017uyz}.}.
The model has vertices as shown in Figure \ref{vertices}.
The corresponding Lagrangian density is (green for $X$, red for $Y$ and yellow for $Z$)
\bea
{\cal L}&=&{\rm Tr}\left(\partial_\mu X \partial^\mu X^\dagger +\partial_\mu Y \partial^\mu Y^\dagger
+\partial_\mu Z \partial^\mu Z^\dagger-m^2 XX^\dagger-m^2 YY^\dagger-m^2 ZZ^\dagger \right.\cr
&&\qquad\left.+g_1 X Y^\dagger Z^\dagger +g_2 X^\dagger YZ\right)
\eea
The Lagrangian is not hermitian.
We have used different couplings for the two different vertices. 
The theory appears to be non-perturbatively unstable, but we will study it in perturbation theory around the point
$X=Y=Z=0$.
In the massless limit it defines a solvable conformal field theory, realizing Zamalodchikov's integrable 
honeycomb \cite{Zamolodchikov:1980mb}.
Specifically, we study the planar limit obtained by taking $N\to\infty$ holding $\lambda_1=g_1\sqrt{N}$ and $\lambda_2=g_2\sqrt{N}$ fixed.

In the planar limit the two point functions $\langle {\rm Tr}(X(x_1)X^\dagger(x_2))\rangle$, $\langle {\rm Tr}(Y(x_1)Y^\dagger(x_2))\rangle$ and $\langle {\rm Tr}(Z(x_1)Z^\dagger(x_2))\rangle$ are not corrected, so we can use the free field two point functions. 
Further the three point functions $\langle {\rm Tr}(X(x_1) Y^\dagger(x_2) Z^\dagger(x_3))\rangle$ and $\langle {\rm Tr}(X^\dagger (x_1) Y(x_2) Z(x_3))\rangle$ are not corrected, so that the $\beta$ functions for these couplings vanish at large $N$. 
Finally, a simplification as compared to the fishnet model is that, for the theory with gauge group $SU(N)$, there are no double trace interactions coming from quantum corrections.

\begin{figure}[ht]%
\begin{center}
\includegraphics[width=1.0\columnwidth]{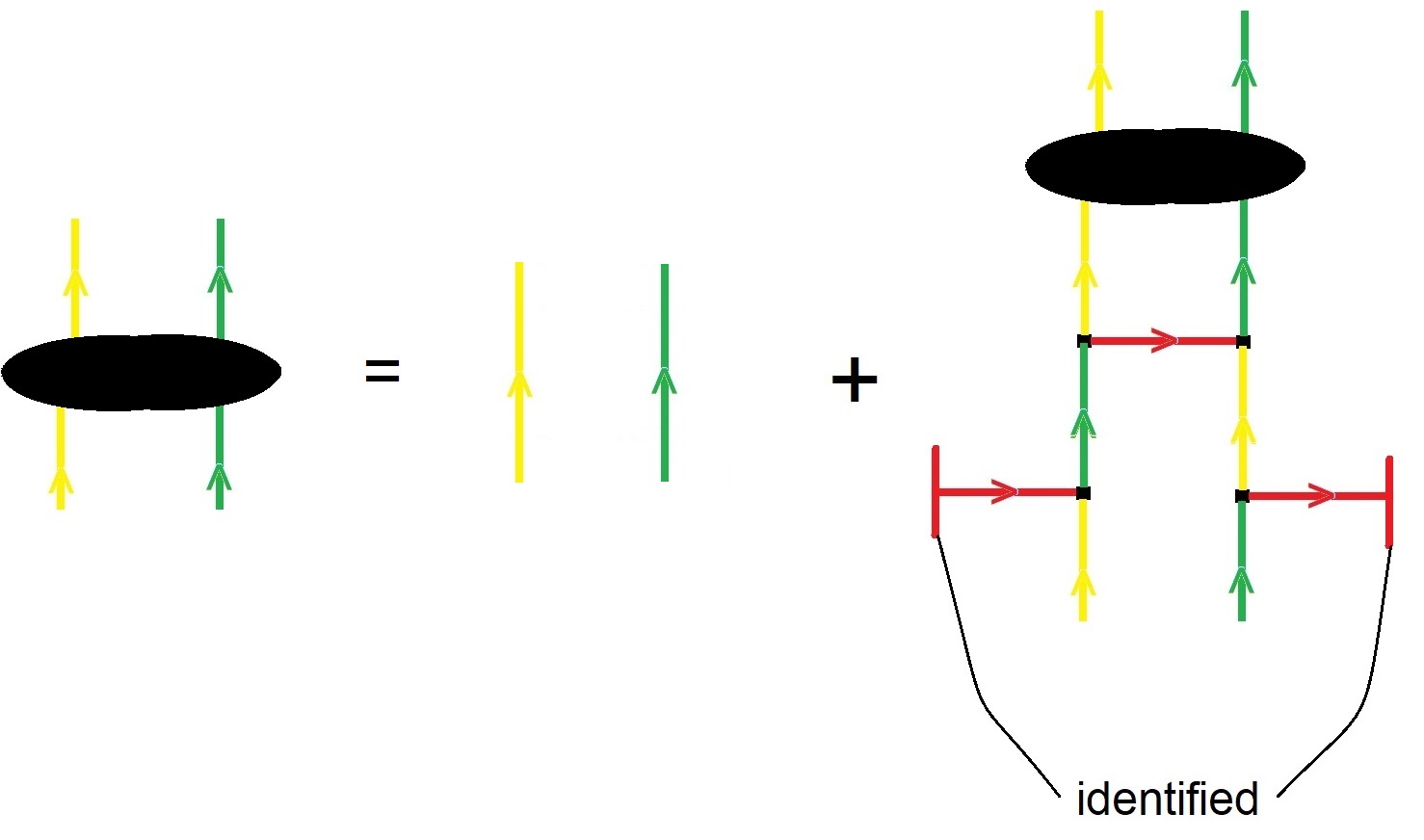}%
\caption{Diagrammatic representation of the integral equation used to define the four point function in the honeycomb 
theory.}%
\label{hexagons}
\end{center}
\end{figure}

The planar four point function is again given by summing ladder diagrams.
The sum of ladders (again denoted $f(\omega,p)$) obeys the following equation 
\bea 
f(\omega,p)=-G_R(p)G_R(\omega-p)\left[1+\int {d^6 k\over (2\pi)^6}R(\omega,p,k)f(\omega,k)\right]\label{SmLd}
\eea
For the honeycomb theory the rung function is
\bea
 R(\omega,p,k)= \lambda_1^2\lambda_2^2 \int {d^6 l\over (2\pi)^6}G_R(l)G_R(l-\omega)
\tilde G(l-p)\tilde G(l-k)\label{hexrung}
\eea
Notice that, in contrast to the fishnet problem, the rung function here does depend on both $\omega$ and $p$.
The diagrammatic version of equation (\ref{SmLd}) is given in Figure \ref{hexagons}.

\vfill\eject

The late time behavior is then given by (make use of (\ref{latetime}))
\bea
-i\omega f(\omega,p)={\pi \over \omega_{\vec{p}}}\delta ( (p^0)^2-\omega_{\vec p}^2))
\int {d^6 k\over (2\pi)^6}R(0,p,k)f(\omega,k)
\eea
which can be transformed to time space to give
\bea
{d\over dt}f(t,p)={\pi \over \omega_{\vec{p}}}\delta ( (p^0)^2-\omega_{\vec p}^2))
\int {d^6 k\over (2\pi)^6}R(0,p,k)f(t,k)
\eea
Notice that we have evaluated the rung function at $\omega=0$ which can be done since we only want the leading
late time behavior.
Now, as above, making the ansatz
\bea
  f(t ,\omega) = f(\omega,\vec{p})\delta (p_0^2-\omega_{\vec{p}}^2)
\eea
and Fourier transforming from $\omega$ to $t$, we find
\bea
{d\over dt}f(t,p)=
\int {d^5 k\over (2\pi)^5}{1\over 4\omega_{\vec k}\omega_{\vec{p}}}\left(R(0,p,k_+)+R(0,p,k_-)\right)f(t,k)
\eea
where $k_\pm = (\pm\omega_{\vec k},\vec{k})$.
Here all vectors are 5 dimensional.
Making the ansatz $f(t,p)=e^{\lambda_L t}f(p)$ we now find
\bea
\lambda_L f(p)=\int {d^5 k\over (2\pi)^5}{1\over 4\omega_{\vec k}\omega_{\vec{p}}}\left(R(0,p,k_+)+R(0,p,k_-)\right)f(k)   \label{HoneyCombEigen}
\eea
The largest eigenvalue of this eigenproblem tells us about chaos in the honeycomb model. 

\section{Numerical Results}\label{HexNumbers}

It is possible to do 5 of the 6 integrals over $l$ appearing in the rung function (\ref{hexrung}) analytically.
The final integral must be dealt with numerically.
First, each $\tilde G$ propagator comes with a delta function.
The integral over $l^0$ can be dome immediately using one of the delta functions.
We then choose a coordinate system for the spatial components of $\vec l$ such that
\bea
\vec p=(p,0,0,0,0)\qquad \vec k=(k\cos\theta,k\sin\theta,0,0,0)
\eea
with $\theta$ the angle between $\vec p$ and $\vec k$.
Using the variables ($0\le L_\perp <\infty$, $0\le\phi_1\le \pi$, $0\le\phi_2 <2\pi$)
\bea
l=(l^1,l^2,\sqrt{L_\perp}\sin (\phi_1),\sqrt{L_\perp}\cos (\phi_1)\sin (\phi_2),
\sqrt{L_\perp}\cos (\phi_1)\cos (\phi_2))
\eea
we can immediately, thanks to rotational invariance, do the integrals over $\phi_1$ and $\phi_2$.
The remaining delta function can be used to do the integral over $L_\perp$.
Next, changing variables from $l^1$ and $l^2$ to $v_1$ and $v_2$ where
\bea
l^1 & = & \frac{(p - k \cos\theta) v_2 + k \sin\theta v_1}{|\vec{k} - \vec{p}|} + \frac{1}{2}(p + k \cos\theta) \cr
l^2 & = & \frac{(p - k \cos\theta) v_1 - k \sin\theta v_2}{|\vec{k} - \vec{p}|} + \frac{1}{2}k\sin\theta
\eea
we are able to do the integral over $v_1$ with the residue theorem.
Thus, in the end we are left with a single integral to do
\bea
R(0,p,k)&=&\int_{-\infty}^{\infty} \theta(f(v_2)) d v_2 \frac{1}{|k^0-p^0| 
\sinh( \frac{\beta}{4}|k^0-p^0 - \frac{2 v_2 |\vec{p} - \vec{k}|}{k^0-p^0}|)
\sinh( \frac{\beta}{4}|k^0-p^0 + \frac{2 v_2 |\vec{p} - \vec{k}|}{k^0-p^0}|)}\cr \cr
& & \times \frac{|\vec{k} - \vec{p}|^2}{32\pi^2 k^2 p^2 \sin^2(\theta)} \left(
\frac{|F(v_2)|}{\sqrt{F(v_2)^2 - f(v_2)}}-1 
\right) \label{RungFinal}
\eea
where
\bea
f(v_2) & = & \frac{1}{4}\left(|\vec{k}-\vec{p}|^2-(k^0-p^0)^2\right)\left(\frac{4v_2^2}{(k^0-p^0)^2}-1\right)-m^2
\eea  
and
\bea
F(v_2) = \frac{k p(k^0 + p^0)\cos\theta - p^2 k^0 - k^2 p^0}{kp(k^0 - p^0) \sin\theta} v_2 + \frac{|\vec{k} - \vec{p}|(k^0 p^0 - k p \cos\theta)}{2 k p \sin\theta}
\eea
The final integral is evaluated approximately. 
First, rescale the $v_2$ coordinate as follows
\bea
v_2 = \frac{a}{2 |\vec{p} - \vec{k}|} V_2.  
\eea
Then, in the integrand make the replacement
\bea
\sinh( \frac{\beta}{4}|k^0-p^0-V_2|)\sinh(\frac{\beta}{4}|k^0-p^0+V_2|)\rightarrow 
\frac{\beta^2}{16} \left| \frac{8}{\beta^2}\left(\cosh\left(\frac{(k^0-p^0)\beta}{2}\right) - 1\right) - V_2^2\right|\cr\cr
\label{JacosApprox}
\eea
This replacement is accurate for small $\beta$ and has the advantage that now the final integral of $V_2$ can be
performed analytically.
Further, due to the $\sinh$'s in the denominator of (\ref{RungFinal}) the large momentum contribution is exponentially
suppressed.  
The $\cosh$ on the RHS of (\ref{JacosApprox}) ensures this suppression, which becomes an issue when the eigenvalue
problem for $\lambda_L$ is solved.
We have verified that this approximation is accurate at small $\beta$ by comparing with a direct numerical integration
of (\ref{RungFinal}).
Denote this small $\beta$ rung function by $R_\beta (p,k,\theta)$
The eigenvalue problem we need to solve is
\bea
\lambda_L f(p)=\int\frac{d^5 k}{(2\pi)^5}
\frac{R_\beta(p,k,\theta)}{4\omega_{\vec{k}}\omega_{\vec{p}} }  f(k)  
\eea
Thanks to rotational invariance, three of the integrals can be done analytically, leaving an integral over the magnitude
$|\vec k|=k$ and an integral over the angle $\theta$ between $\vec{k}$ and $\vec{p}$
\bea
\lambda_L (p^2 f(p)) = 2\pi^2 \int_0^\infty dk \left( \int_0^\pi d\theta\,\, \frac{p^2 k^2 \sin^3\theta}{(2\pi)^5}\,
\frac{R_\beta(p,k,\theta)}{4\omega_{\vec{k}}\omega_{\vec{p}}}
 \right) (k^2 f(k))
\eea
After discretizing, the term in brackets defines a matrix whose eigenvalues give the growth exponent $\lambda_L$.
We consider eigenfunction $p^2 f(p)$ so that discretization gives a symmetric matrix that can be diagonalized.
As before we make the transformation (\ref{transk}), which maps the infinite $k$ interval onto the interval $[0, 1]$.  
The transformation has a Jacobian of $\frac{3}{(1 - u)^2}$.  
Discretize the $u$ and $v$ intervals uniformly with spacing $\delta u$ to obtain the matrix
($k_n=k(u_n)$ and $p_m=p(v_m)$)
\bea
R_{mn} = {\delta u\over 64\pi^3} \left( \int_0^\pi d\theta 
\frac{3k_n^2 p_m^2 \sin^3\theta}{(u-1)(v-1)\omega_{\vec k_n}\omega_{\vec p_m}}
R_\beta(k_n,p_m,\theta)\right)
\eea
whose largest eigenvalue gives the growth exponent $\lambda_L$.  We have again absorbed a factor into the eigenfunction
to ensure a symmetric $R_{mn}$. 

\begin{figure}[ht]%
\begin{center}
\includegraphics[width=0.7\columnwidth]{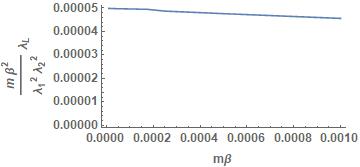}%
\caption{Behavior of the growth exponent $\lambda_L$ for the honeycomb model.
}%
\label{Hexgrowthexponent}
\end{center}
\end{figure}

Similar to the case of the fishnet model, we find that $R_{mn}$ only has positive eigenvalues.
This is presumably again because there is no correction to the two point function, implying that even at finite
temperature our single particle states have an infinite lifetime.
The results for the growth exponent, $\lambda_L$, at small $\beta$ are given in Figure \ref{Hexgrowthexponent}.

\vfill\eject

\section{Discussion}

The fishnet CFT is a fascinating model. 
It is an interacting non-supersymmetric CFT in four dimensions, which is rare and already makes it interesting.
Further, it exhibits integrability which it inherits from its ${\cal N}=4$ super Yang-Mills theory parent.
For integrability to work out, conformal invariance is crucial.
In this study we have used the same model, but with a mass deformation which obviously ruins conformal invariance.
The resulting model is still a beautiful model with enough simplicity that we can solve for correlation functions exactly
 in the planar limit.
In this way, the model has the potential to teach us about phenomena that may be hard to reach at weak coupling.

We have studied certain out of time order four point functions and have found that the model exhibits chaos.
Further, it violates the bound of \cite{Maldacena:2015waa}.
Our model is not unitary and we interpret the violation of the chaos bound as evidence that unitarity is an important
ingredient in the bound of \cite{Maldacena:2015waa}.
The fact that we have managed to carry this computation out in the fishnet model suggests it is a good model to keep in 
mind for other computations.
An obvious question is to ask if the spectral form factor can be computed in this model.

One could also ask how the planar results will get corrected once non-planar corrections are included.
It is clear that there will be non-planar corrections to the two point functions, and that at two loops, these will
have an imaginary part.
Thus, a finite lifetime for particles is a non-planar effect in the fishnet (and honeycomb) models.
Once included, we expect an exponential decay of the two point functions and hence thermalization.

Finally, we could repeat the computations we have carried out in this study for the three dimensional sextic model too.
Based on the results of this study, we expect the three dimensional model exhibits chaos and that again the chaos bound 
will be violated.

{\vskip 0.5cm}

\noindent
\begin{centerline} 
{\bf Acknowledgements}
\end{centerline} 

We would like to thank Volodya Kazakov for correcting some errors in the first version of this article. 
This work is supported by the South African Research Chairs
Initiative of the Department of Science and Technology and National Research Foundation
as well as funds received from the National Institute for Theoretical Physics (NITheP).

\end{document}